# Nucleation and Antiphase Twin Control in $Bi_2Se_3$ via Step-Terminated $Al_2O_3$ Substrates

Alessandro R. Mazza[1#], Jia Shi[2], Gabriel A. Vázquez-Lizardi[3], Sangsoo Kim[4], Jackson Bentley[4,5], An-Hsi Chen[4], Kim Kisslinger[6], Debarghya Mallick[4], Qiangsheng Lu[4], T. Zac Ward[7], Vitalii Starchenko[8], Nicholas Cucciniello[9,10], Robert G. Moore[4], Gyula Eres[4], Yue Cao[11], Debangshu Mukherjee[12], Liam Collins[7], Christopher Nelson[7], Danielle Reifsnyder Hickey[3,13,14], Fei Xue[2], Matthew Brahlek[4*]

[1]Materials Science and Technology Division, Los Alamos National Laboratory, Los Alamos, NM 87545, USA
[2]Department of Physics, University of Alabama at Birmingham, Birmingham, AL, 35294, USA
[3]Department of Chemistry, Pennsylvania State University, University Park, PA 16802, USA
[4]Materials Science and Technology Division, Oak Ridge National Laboratory, Oak Ridge, TN, 37831, USA
[5]Department of Physics and Astronomy, Vanderbilt University, Nashville, TN, 37240, USA.
[6]Center for Functional Nanomaterials, Brookhaven National Laboratory, Upton, New York 11973, USA
[7]Center for Nanophase Materials Science, Oak Ridge National Laboratory, Oak Ridge, TN, 37831, USA
[8]Chemical Sciences Directorate, Oak Ridge National Laboratory, Oak Ridge, TN, 37831, USA
[9]Center for Integrated Nanotechnologies, Los Alamos National Laboratory, Los Alamos, NM 87545, USA
[10]Department of Materials Design and Innovation, University at Buffalo – The State University of New York, Buffalo, New York 14260, USA
[11]Materials Science Division, Argonne National Laboratory, Lemont, IL 60439, USA
[12]Computing and Computational Sciences Directorate, Oak Ridge National Laboratory, Oak Ridge, TN, 37831, USA
[13]Department of Materials Science and Engineering, Pennsylvania State University, University Park, PA 16802, USA
[14]Materials Research Institute, Pennsylvania State University, University Park, PA 16802, USA

Correspondence should be addressed to #armazza@lanl.gov, *brahlekm@ornl.gov

**Abstract**: The epitaxial synthesis of high-quality 2D layered materials is an essential driver of both fundamental physics studies and technological applications. $Bi_2Se_3$, a prototypical 2D layered topological insulator, is sensitive to defects imparted during the growth, either thermodynamically or due to the film-substrate interaction. In this study, we demonstrate that step-terminated $Al_2O_3$ substrates with a high miscut angle (3°) can effectively suppress a particular hard-to-mitigate defect, the antiphase twin. Systematic investigations across a range of growth temperatures and substrate miscut angles confirm that atomic step edges act as preferential nucleation sites, stabilizing a single twin domain. First-principles calculations suggest that there is a significant energy barrier for twin boundary formation at step edges, supporting the experimental observations. Detailed structural characterization indicates that this twin-selectivity is lost through the mechanism of the 2D layers overgrowing the step edges, leading to higher twin density as the thickness increases. These findings highlight the complex energy landscape unique to 2D materials that is driven by the interplay between substrate properties, nucleation dynamics, and defect formation, and overcoming and controlling these are critical to improve material quality for quantum and electronic applications.

## Introduction

Two-dimensional (2D) materials have risen to prominence over the past three decades due to the diversity of electronic and magnetic properties as well as the low-dimensional crystal structures which enable exfoliation and assembly of units as thin as a single atomic layer. This advance has coincided with the rise of the topological description of electronic properties of materials, which is a foundation for understanding phenomena that strongly deviates from the simple model of their electrons behaving as semiclassical free particles. This is captured in the novel Dirac- and Weyl-like energy-momentum dispersions of materials ranging from graphene to topological insulators[1] such as the $Bi_2Se_3$ tetradymite family[2–4] to the zoo of topological semimetals[5] , ranging from 3D systems such as $Na_3Bi$[6] to $Cd_3As_2$[7,8] and Weyl systems such as TaAs[9,10], as well as the recent discoveries of their 2D analogs[11,12]. A central theme that has enabled rapid progress is the synergy between materials synthesis and fundamental physics, which is driven by the growth of bulk[13] and thin film materials[14] as well as the process of exfoliating and assembling those materials to discover and benchmark new physics[15]. Specific to the thin film growth of 2D materials are challenges concerning defects that arise due to weak interlayer coupling. Moreover, topological materials present additional challenges since their structures are simultaneously extremely sensitive to charge and prone to the formation of structural defects, both of which are tied chemically to the same properties that make them good topological materials[16]. As such, elucidating the basic science of how these materials grow as epitaxial thin films can permit the production of higher-quality materials that exhibit intrinsic properties, and, furthermore, enable mechanisms to realize new platforms for designing and stabilizing new physical phenomena.

The focus of this paper is on the critical parameters for the growth of layered topological insulators by molecular beam epitaxy (MBE), which are applicable to 2D materials growth in general. $Bi_2Se_3$, the focus of this study, has become the prototypical topological insulator due to its large band gap of 0.3 eV and simple band structure with a single Dirac cone that is centered between the conduction and valence bands. Also, the defect chemistry of $Bi_2Se_3$ is very simple, as the primary defect in MBE-grown films are Se-vacancies[4]. This contrasts with $Bi_2Te_3$ and $Sb_2Te_3$, which have the Dirac point much closer to their valence bands and have more complex defect chemistries due to the size and chemical similarities of Bi, Sb, and Te[4]. However, these more complex defect chemistries of $Bi_2Te_3$ and $Sb_2Te_3$ have enabled charge compensation doping schemes to be successful by cation alloying of Bi and Sb and anion alloying of Se and Te[4,17,18].

Thin films of all the aforementioned materials – $Bi_2Se_3$, $Bi_2Te_3$ and $Sb_2Te_3$ – suffer from additional defects due to the substrate/film interface, such as chemical reactions and structural aspects such as lattice relaxation resulting from large lattice mismatches[4,14,19–23]. Since these materials are composed of layers with triangular lattices, they are additionally prone to antiphase twins (APT) where the nucleated layers are aligned with respect to the substrate in one of the two equivalent orientations. Earlier work to mitigate twinning in $Bi_2Se_3$ films has focused on utilizing InP(111), which has a close lattice match, and has resulted in films that were closer to a single domain[24–26]. However, for InP(111), the interfacial structures were found to be disordered, and the low twin-density might be tied to the intrinsic roughness of the substrate prior to growth[26] compared to flat surfaces of InP[27]. There has been success for the growth of $Bi_2Se_3$ on the isostructural 2D polymorph of $In_2Se_3$ and $(Bi_{0.5}In_{0.5})_2Se_3$, either from cleaved crystals[28,29] or epitaxial buffer layers[20]. This has substantially increased electron mobility[20,30] as well as being shown to be effective at reducing twins[31]. However, In up-diffusion has been noted to be a problem since a small amount of In can drive the film non-topological[32–35], and these materials are not widely available as wafer-scale substrates. These issues demonstrate that understanding how to synthesize $Bi_2Se_3$ and 2D materials, more generally, on a widely available substrate with control over defect formation is critical for understanding new physics and pushing toward new applications.

Here, we show how step-terminated $Al_2O_3$ substrates can be used to optimize single-domain $Bi_2Se_3$ growths. This is achieved using $Al_2O_3$ with a high, 3° miscut, which is a critical advancement for this material. Moreover, we elucidate the critical parameters that affect the growth of $Bi_2Se_3$, systematically comparing films grown over a range of temperatures and miscut angles. Varying the terrace width reveals that step edges act as nucleation sites. This indicates that sufficient thermal energy is needed to increase the mean free path of absorbed atoms to ensure interaction with the substrate step edges. First principles calculations reveal that there is a large disparity in total energy for twin boundaries to form on the steps compared to untwinned, which suggests that the local bonding to a step edge in $Al_2O_3$ is the key factor that drives this selection process and explains previous results. Detailed structural characterization reveals that the step-edge phase selection is lost through a novel mechanism that is unique to 2D materials. Specifically, it is found that the 2D layers of $Bi_2Se_3$ can overgrow the step defects, thus eliminating them as nucleation points and thereby the phase selection is generally lost for thicker samples. These aspects together highlight many unique features for the growth of 2D materials like $Bi_2Se_3$, motivating additional studies (both theory and experiment), which are necessary for overcoming many quality-related barriers. More specifically, our work gives insight into how single-domain films can be grown on substrates with the smallest miscut, and hence minimal defects, which are critical to applications and a better understanding of the unique properties in this material class.

**Results**

Figure 1(a) shows schematic structures of both $Al_2O_3$ and $Bi_2Se_3$ on the left and right, respectively. The top images are projections perpendicular to the b-axis and the bottom are the projections perpendicular to the c-axis. Several important aspects related to epitaxial growth can be seen. First, the unit cell of $Al_2O_3$ is composed of oxygen octahedral units that are around 2 Å in height. These units are assembled to give a unit cell that is 12.99 Å along the c-axis. For $Bi_2Se_3$, the structure is composed of Se-Bi-Se-Bi-Se units. These 5-atom layers are called quintuple layers (QLs), which are van der Waals bonded along the c-axis and are around 10.0 Å thick. Both $Al_2O_3$ and $Bi_2Se_3$ are triangular in the a-b plane, as shown in the lower panel of the figure. This highlights the correct symmetry match for epitaxial compatibility and growth. However, the lattice mismatch is quite significant. Specifically, the strain-mismatch for $Bi_2Se_3$ ($a = b = 4.14$ Å) relative to $Al_2O_3$ ($a = b = 4.75$ Å) is around 15%. Despite this large mismatch, $Bi_2Se_3$ grows well on $Al_2O_3$ with the relaxation occurring fully in the first QL[14,23,36], owing to the weak out-of-plane van der Waals bonding[37]. However, there is much unknown regarding how this relaxation occurs and there are implications on the physical properties such as large defect densities confined to this interface. Other substrates with better lattice matching such as InP(111) have been used, as mentioned above, but the tradeoff for better lattice match is higher chemical reactivity. In contrast, $Al_2O_3$ is relatively inert, easy to prepare, cheap and available in high-quality wafers.

A comparison of the structures of $Al_2O_3$ and $Bi_2Se_3$ reveals another challenge. The structures have similar triangular in-plane surfaces, which are highlighted by the black coordinates for $Al_2O_3$ and the red coordinates for $Bi_2Se_3$. On the $Al_2O_3$ surface two $Bi_2Se_3$ coordinates are overlaid, one with solid lines and a second that is rotated by 60°, which are labeled as *T* and *T'*, respectively. By inspecting the structural alignment of $Bi_2Se_3$ with *T* and *T'* phases on $Al_2O_3$, one can see that both orientations are equivalently matched to the $Al_2O_3$ structure below. This implies that, from a structural perspective, there should be APT of *T* and *T'* and they should be in equal proportion. To quantify the relative stability of $Bi_2Se_3$ on $Al_2O_3$, we performed density functional theory (DFT) calculations of the total energy for two registry variants ($T$ and $T'$)[38–40]. All simulations were carried out in VASP using the projector–augmented–wave (PAW) method with the Perdew–Burke–Ernzerhof (PBE) functional within the generalized gradient approximation (GGA)[41]. We used a 400 eV plane-wave cutoff (converged to 1 meV/atom and residual force less than 0.01 eV/ Å$^{-1}$) and constructed slab models from experimental $Al_2O_3$ and $Bi_2Se_3$ unit cells to minimize lattice mismatch. Structural relaxations employed a Monkhorst–Pack k-point mesh of 1×1×1 (Fig. 1) and 1×3×1 (Fig. 2)[42]. The simulated super cells are shown in Fig. 1(b) on the left and right, respectively. Here, a single QL of $Bi_2Se_3$ that spans several in-plane pristine unit cells was used to account for the large lattice mismatch. After relaxation, both the $T$ and $T'$ configurations converged to geometries closely matching their bulk phases. The total energies are essentially identical: −1680.0459 eV ($T$) and −1680.0461 eV ($T'$), a supercell difference of ~0.138 meV (≈0.7 μeV per atom; both at −5.90 eV/atom). This minute difference lies well within the numerical uncertainty associated with interfacial registry (e.g., the relative alignment of $Bi_2Se_3$ on $Al_2O_3$) and standard DFT settings (e.g., choice of pseudopotentials), and is negligible compared with thermal energies at typical synthesis temperatures (200–300 °C). We therefore regard the $T$ and $T'$ registries on $Al_2O_3$ as effectively energetically degenerate.

To achieve films with a single twin the energy degeneracy of $Bi_2Se_3$ on the $Al_2O_3$ surface must be broken. $Al_2O_3$ with a c-plane surface is well-known to exhibit atomic steps when annealed to ~1000 °C. Here, $Al_2O_3$ will reconstruct, and the steps are formed by aluminum-oxygen octahedral units with a height of around 2 Å, as shown in Fig. 2(a). In this scenario, the step edges can act as nucleation sites that may stabilize one twin orientation over another. In Fig. 2, DFT stabilized structures for $Bi_2Se_3$ at the 2 Å steps were relaxed, and their total energies were compared. Untwinned structures with $Bi_2Se_3$ layers in the $T$-$T$ (b) and $T'$-$T'$ (c) are compared to twinned structures in $T'$-$T$ (d) and $T$-$T'$ (e). As before, the structures are composed of multiple $Bi_2Se_3$ cells in-plane as well as out-of-plane. However, as these are very large calculations, the atoms away from the steps were fixed to their bulk-like structures. Near the step-edge and all Bi, Se, Al, and O atoms were allowed to relax to their lowest energy configuration. The $T$–$T$ structure has the lowest total energy (≈ –5.97 eV/atom), so we reference all other configurations to it. The $T'$–$T'$ structure is 10 meV/atom higher in energy than $T$–$T$, which is not the case in the flat interface (Fig. 1(b)) where $T$ and $T'$ are almost energetically degenerate. The twinned structures were found to be significantly higher in energy with $T'$-$T$ being 80 meV/atom higher in energy and $T$-$T'$ being 100 meV/atom. Since the calculations were taken with the "bulk" of the layers fixed in positions, the difference in energy comes primarily from the atoms near the steps. Thus, the energy difference might, in fact, translate to a large total energy difference between the different configurations. The energy difference is of the order or larger in comparison to the thermal energy scales used at growth temperatures typical for $Bi_2Se_3$, ~200-300 °C (here 200 °C ≈ 40 meV), which suggests that growing $Bi_2Se_3$ on step-terminated $Al_2O_3$ may provide a route to stabilize films with minimal APT.

To test this hypothesis, we used MBE to grow $Bi_2Se_3$ on step-terminated $Al_2O_3$ substrates which had miscuts of <0.05°, 0.1°, and 0.2° (purchased from Shinkosha Co., LTD). Substrates with 3° miscut were prepared by annealing epi-ready substrates to 1000 °C in air for one hour[43]. Prior to growth the substrates were mounted to the sample platens using Ag-paste and cured under vacuum for around 10 minutes at 150 °C. The substrates were exposed to UV-ozone for 10 minutes prior to being placed into the MBE system and transferred into the main reactor. The substrates were then heated to 600 °C, while being exposed to Se, then cooled down to the growth temperature. The growth rate was around 1 QL per minute and the Se flux was ~10× the Bi flux. We chose two growth methods for comparison. The first is the Rutgers' method for $Bi_2Se_3$ which splits the growth into two steps[14,23]. The first step is a low-temperature (150 °C) growth of 3 QL. The low temperature promotes nucleation and ensures this layer is conformal in the first QL. The films are initially crystalline but of low quality during this step. However, once the substrate is heated to 235 °C, the subsequent growth occurs with significantly improved crystallinity until the target thickness is achieved. The second method employed here is a single-step growth. For nucleation to be controlled by the step edges the adsorbed atoms (adatoms) should have sufficiently large mean-free-paths on the surface to diffuse and interact with the atomic steps. The single-step growth enabled two parameters to be varied, the temperature and the miscut.

The crystallinity was probed using reflection high-energy electron diffraction (RHEED) and X-ray diffraction. RHEED images along the [110] and [100], shown in Fig. 2(a), exhibited bright, well-localized diffraction peaks and clear streaks, which indicate that the surfaces were 2D and highly crystalline. The film shown was grown using a single step at 280 °C to 20 QL, and there were no noticeable differences for films grown under different conditions since the beam will see the same periodicity for *T* and *T*' (see Fig. 1(a)). Figure 3(b) shows specular X-ray diffraction for $Bi_2Se_3$ films grown on step-terminated $Al_2O_3$. The measurements were done on either a 4-circle Panalytical MRD using Cu $K_{\alpha 1}$ radiation or a 4-circle Rigaku SmartLab also monochromated to Cu $K_{\alpha 1}$. The data shown in Fig. 3(b) are for a sample grown using the 2-step (thickness ~15 QL) method on a substrate with miscut that was <0.05° as well as single-step growths at a temperature of 280 °C where the substrate miscut values were <0.05°, 0.1°, and 0.2° (thickness ~12 QL). All data show similar features. First, the main 00*L* $Bi_2Se_3$ reflections can be seen for $L$ = 3, 6,…, 21, which reflects the *a*-*b*-*c* stacking QL periodicity. This yields a QL thickness very close to the 10.0 Å bulk value. About these main peaks are clear oscillations which arise from coherent interference between x-rays scattered off the top surface and the interface with $Al_2O_3$. These indicate that the film's crystalline surfaces are all structurally coherent along with the entire film's thickness. As expected, the 2-step sample has more pronounced Laue oscillations, which are likely more prominent since the films are flatter due to the low temperature step which promotes conformal coverage. Finally, for each single-step growth there are two curves shown, which differ by an offset in $\theta$. The solid-colored curves were obtained by aligning $\theta$ with respect to the $Bi_2Se_3$ 006 and the dashed curves were obtained by aligning to the $Al_2O_3$ 006. Using this procedure resulted in an imperfect alignment to the substrate step direction and therefore, there was no clear relation among the miscuts and the offsets of the two alignments. If the 00*L* planes of $Al_2O_3$ and $Bi_2Se_3$ were perfectly aligned, then the curves would overlap, as is found for the samples grown on the substrates with miscut <0.05°. However, for the 0.2° and 0.1° miscuts, this was found not to be the case. This implies that the 00*L* planes of the $Bi_2Se_3$ are likely aligned with respect to the surface of $Al_2O_3$ rather than with the crystallographic direction of $Al_2O_3$. How this occurs from a microstructural perspective is unknown and remains a point for further study, though this nonconformal growth has been observed in other 2D systems such as graphene on SiC[44]. It is noted that the rocking curve width for these films were around 0.020° for the $Bi_2Se_3$ compared to 0.004° for the $Al_2O_3$, which is consistent with reported values[24,27]. Altogether, the RHEED and X-ray diffraction indicates that the films are of very high structural quality as grown with both 1-step and 2-step growth methods.

To quantify the APT density, the off-axis 105 peak was measured as a function of azimuthal angle, $\phi$. The 105 peak was chosen since it should have 3-fold multiplicity for an untwinned sample, whereas a twinned sample will exhibit 6-fold periodicity. Figure 4 summarizes the data for samples grown at different temperatures as well as miscuts. Figure 4(a) shows azimuthal scans for the same set of samples as shown in Fig. 3, where the most intense peak is labeled as *T* and the less intense peak labeled with *T'*. The data were all normalized to the most intense peak. It is noted that the $Al_2O_3$ 104 peak coaligned with the *T* orientation. For the 2-step sample, the short mean-free path is very small during the initial low temperature growth seeding both *T* and *T'* with nominally the same portions. In contrast, the single-step samples show greatly reduced proportion of *T'* in comparison to *T*, which confirms the effectiveness of step-edges in reducing APTs.

To better understand the growth properties, Fig. 4(b-c) shows the APT fraction as a function of both growth temperature (b) and miscut (c). The vertical axis is the APT fraction which is defined as the intensity of *T'* divided by the sum of the intensity of *T* and *T'*, hence a twinned film will be 50% (horizontal dashed line) and an untwinned sample will be zero. With this metric, the 2-step sample is close to 50%, as indicated. In contrast, the APT fractions for the single-step growth all monotonically decrease with increasing temperature. The APT fraction is systematically reduced with increasing miscut. (It is noted that the actual sample temperature uncertainty is estimated to be as high as ~±10-15 °C sample-to-sample, shown as horizontal error bars, and that maintenance performed on the MBE reactor may have introduced a systematic offset among the data in Fig. 4(b) and 4(c).) To control nucleation, the adatoms need to have a high probability of interacting with the steps. Thus, the mean-free-path on the surface needs to be of the order of the step separation (terrace width) and thus should show strong dependence on both the temperature as well as the miscut, which is in direct agreement with the data shown in Fig. 4(b).

In Fig. 4(b), for the case of a miscut of 0.2° and a growth temperature of 280 °C, the APT fraction reaches a minimum value of around 15%. For these miscuts and film thickness, this might represent an upper limit since $Bi_2Se_3$ starts evaporating at temperatures above 280 °C, as indicated in the figure. To push the APT fraction down, the terrace width should be reduced. As such, the temperature was fixed at 235 °C, and a series of samples were grown with substrate miscut ranging from <0.05° up to 3°. Assuming single 2 Å, steps the miscut angle geometrically translates to a step-to-step distance, or terrace width, ranging from several hundred nanometers for the 0.05° miscut to around 5 nm for the 3° miscut ($w = a \times \tan(90-\delta)$ where $w$ is the terrace width, $a$ is the step height, and $\delta$ is the miscut). As shown in Fig. 4(c), the APT fraction reduces monotonically with increasing miscut (reducing terrace width). For the 3° miscut, the *T'* APT is undetectable, which indicates the growth is entirely controlled by the nucleation at the step edges. This experimentally suggests that the steps of $Al_2O_3$ can preferentially nucleate single APT films. To interrogate the crystalline microstructures, Fig. 5 shows high-angle annular dark-field scanning transmission electron microscopy (HAADF STEM) images of the 12 QL sample grown at 280 °C on 0.2° miscut. The specimens for HAADT STEM measurements were prepared using focused ion beam (FIB) lift-out techniques on a FEI Helios G5 UX DualBeam FIB/SEM system, with final $Ga^+$ milling performed at 2 keV. STEM was performed at 100 kV on a Nion UltraSTEM. In Fig. 5, the beam direction (in-to-the-page) is along the [010] crystallographic direction. In Fig. 5(a), the QL structure is clearly visible and homogeneous across this wide-scale image, which overall agrees with the macroscale x-ray diffraction analysis. Fig. 5(b) shows a zoomed-in image from the boxed region in Fig. 5(a). Here the QL stacking is visible as well as additional microstructural information detailing the APT and interfacial properties. In this image, the majority of the film is a *T* domain. However, at the top of the film there is a small portion of the *T'* orientation, as highlighted by the yellow horizontal arrows. Similarly, there is a portion of the film in the image on the bottom right where the QL structure is distorted at a *T* and *T'* boundary, with associated lattice disorder that is causing a strain gradient and subsequent distortion, consistent with previous observations of in-plane rotations[22]. Moreover, there exists a distinct interfacial region between the $Bi_2Se_3$ and the $Al_2O_3$. This region is more easily seen in the magnified image shown in Fig. 5(c) and is not a disordered QL nor a disordered layer of $Al_2O_3$. The height of this layer is less than a QL at around 0.6-0.7 nm, and the Z-contrast indicates that the density is below that of $Bi_2Se_3$ and larger than $Al_2O_3$. Moreover, while it appears somewhat amorphous, there is a visible 2-atom-thick order within this layer, which can be seen in Fig. 5(c).

Likely, this layer is semi-crystalline and is roughly aligned crystallographically with the $Al_2O_3$, which facilitates epitaxial alignment of $Bi_2Se_3$ with $Al_2O_3$. This indicates that it is likely a Se-rich passivation layer of the $Al_2O_3$ or perhaps BiSe, which has been observed at interfaces with $NbSe_2$[45]. It is noted that right below this disordered layer there are atoms that have clearly diffused down into the $Al_2O_3$ layer. Due to their brightness, the atomic columns must contain elements heavier than Al and O and thus are likely Se occupying the ordered vacancies in the $Al_2O_3$.

Finally, in Fig. 5(d), we show a high-magnification image of one of the defects found in these films that is associated with the step edges. The defect clearly emanates from the substrate which has a 2 Å step. In the lower portion of the image there is a vertical twin boundary that has a nominal 2 Å offset between the layers along the c direction. This defect is comparable to the calculated models in Fig. 2 (*T-T’* domain) showing very clear bonding across the QL units. This specific defect was chosen since it is one of the clearest instances we found regarding how these defects terminate: At the top of the image the defect vanishes by a process of 2D layer overgrowth, which is a critical mechanism to understand for stabilizing a single APT. Within this continuous layer there are clear areas of shear distortion which are most likely a strain gradient that arises due to the vertical offset from the $Al_2O_3$ step. Much like a carpet being draped over a staircase, the 2D QL units overlay the edge defect. This process thus exchanges the highly disordered bonding environment of the defect for a continuous, albeit shear strained, layer on top. This mechanism of exchanging the energy cost for a defect with the energy cost of the strained layer is probabilistic, and thus the density of step edges should decrease as the film thickness increases. Moreover, the irreversibility of this transition is critical for understanding the fundamentals of the growth and the ability to stabilize films with a single APT, since a region with a strain gradient will not act as a nucleation center like a step-edge defect site. More precisely, the defect associated with steps clearly favors nucleation of a single twin, yet as films grow thicker it is observed that they tend towards equal domains, which is most likely driven by the passivation of nucleation sites by the 2D layers covering the defect as highlighted in Fig. 5(d).

To better understand the role of substrate steps in growth, Figure 6 shows atomic force microscopy (AFM) images of a 5 QL (a-c) and 20 QL (d-f) films grown at 260 °C with a single step process on 0.05° (a, d), 0.1° (b, e) and 0.2° (c, f) miscut. The measurements were performed using a DriveAFM system (Nanosurf AG, Switzerland) operated in WaveMode. A setpoint force of 5 nN was maintained during imaging to ensure stable tip–sample contact while minimizing tip-induced deformation. We employed NSC36 silicon cantilevers (MikroMasch) with resonance frequency ~75 kHz, force constant ~2 N/m, and tip radius <10 nm. All AFM data were processed using Gwyddion (version 2.66). To remove background artifacts and scanner bow, each image was first-order flattened prior to further statistical analysis. Roughness parameters including RMS roughness (Sq), mean roughness (Sa), skewness (Ssk), were extracted from the flattened topography maps. Representative images are shown in the main text, and full datasets are available upon request. The dominant feature seen in all images are triangles, which are steps of 1 QL either islands (highlighted as solid triangles in Fig. 6(b)) or holes (dashed triangles in Fig. 6(b); it is noted that the holes are only on the topmost surface and not through to the substrate), whose shape reflects the underlying natural habit of $Bi_2Se_3$ as it grows. Moreover, the 5 QL all show steps from the substrate, which are the sharp parallel lines and appear in AFM since there is no structural mechanism to get rid of these steps during the growth. Overall, the films are extremely flat, only varying by ±1 QL across the whole of the images for the 5 QL and around ±2 QL for the 20 QL films, with RMS roughness around 1-2 nm (see Fig. 6).

Looking closely at these features gives additional insight into the growth mode of the films as they are snapshots of the surfaces when the growth terminates. The orientation of the triangular features reflects the in-plane structure of the surface, which, for clarity, is highlighted in Fig. 6(b) by the white and yellow triangles. These triangular features point in one of two directions, an indication of the existence of both *T* and *T'* phases, as labeled. For the 5 QL samples, the surface morphology changes significantly with miscut. The sample with 0.05° miscut has a larger density of smaller islands (~100 nm) compared to the 0.1° and 0.2° samples where the islands are much larger (~300 nm) and sparser. Moreover, as highlighted in the inset of Fig. 6(a), the 0.05° sample has additional small islands. These results together suggest that the 0.05° sample exhibits a higher degree of random nucleation, consistent with the analysis that in this temperature regime the mean-free-path of the adatoms on the surface is similar to the step-terrace width for 0.05° but sufficiently larger for the samples grown on 0.1° and 0.2°. Regarding the origin of the islands, all samples exhibit nominally similar density of both orientations. The 0.1° sample has slightly more *T* islands than *T'* islands, which is consistent with the other samples. This suggests that the islands may arise from both oriented nucleation at the steps and random nucleation. In comparison, the holes arise where growth fronts merge, which explains the opposite orientation of *T* and *T'* holes relative to the *T* and *T'* islands. Interestingly, in the 0.1° sample, there are only two regions that have *T'* alignment and many regions that have *T*. This suggests that the underlying film is close to a single domain. The shape and habit of the holes compared to the islands is far less well-ordered. Despite nucleation control from the steps, this appearance of disorder can be ascribed to the randomness of merging of growth fronts on a highly mismatched substrate, which is consistent with the rich microstructure seen in electron microscopy.

Finally, comparing the differences in morphology for the 5 and 20 QL samples and different miscuts gives more details into how these 2D materials grow with the atomic scale steps. First, the triangular features are less coherent for the 20 QL samples compared to the 5 QL samples, which indicates the nucleation at the steps is mitigated as the thickness increases as the ratio of *T* and *T'* domains increase. This is consistent with the steps being overgrown, as seen in HAADF-STEM image in Fig. 5(d), which no longer act as nucleation sites. Moreover, the 5 QL sample with 0.05° sample has smaller features compared to the 0.1° and 0.2° samples, and the surfaces for the 20 QL samples become less distinct with miscut, highlighting that there is a clear change in growth mode, as discussed more below.

**Discussion**

Here we have shown that it is possible to systematically create $Bi_2Se_3$ films that are composed of single APT by utilizing step-terminated $Al_2O_3$ substrates. This was achieved by employing substrates with varying the miscut, ranging from <0.05° to much higher at 3°. The latter, with a terrace width of around 5 nm, was necessary for attaining films with no APTs. This is a critical achievement for the synthesis of topological materials and 2D materials in general, and the comprehensive characterization and calculations reveal why this is possible. Moreover, this work reveals clear deviations from the simple models for thin film synthesis, which have clear implications for the wider families of 2D materials. This work highlights critical challenges and opportunities regarding the synthesis of 2D materials in general and, specifically, chalcogenide topological materials. First, all the data presented here point to the fact that the growth must be controlled by nucleation at the step edges. This includes the growth temperature dependence, where the APT fraction systematically decreases with increasing temperature and with increasing miscut (reducing terrace width). This reduction is consistent with the increase of the adatom mean-free-path on the surface relative to the terrace width. The step-edge nucleation can only be controlled if the mean-free-path is of order or larger than the terrace width. However, there are many aspects uncovered here that are critical including the (i) interface, (ii) nucleation and growth modes, and (iii) rich aspects of defects, which we discuss below in more depth. Quantifying and understanding these aspects remain open areas of research.

**(i) Interface**: The reduction of twins due to nucleation at the steps is convoluted with the role of the interfacial layer observed in Fig. 5. The structural and chemical composition of this layer is difficult to determine, but it is likely a Se-rich phase based on the Z-contrast. Critical to achieving single APT and understanding the growth in better detail requires clarifying what this layer is and its properties. First, does this layer form during the initial monolayer of $Bi_2Se_3$, or does it form in part during the high-temperature substrate treatment that was used to in-situ clean the surfaces? Systematically changing treatment steps and analyzing the resulting film quality could give additional insight; however, skipping the cleaning step may convolute surface contamination with the chemically adsorbed interface layer. Furthermore, the role of the surface diffusion properties of Bi and Se adatoms on this layer might be critical. From a chemical perspective, the $Al_2O_3$ may be very sticky in that the surface diffusion coefficient may be relatively short, in contrast to 2D materials such as graphene where the diffusion coefficient is very long due to the lack of bonding sites. As such, this interfacial layer might enhance the surface diffusion, for example, if the chemical bonding is relatively 2D. Similar self-buffered layers have been found in other heteroepitaxial systems where there is a large degree of chemical and epitaxial mismatch and a defective monolayer anomalously forms between the film and substrate[46,47], which might point to a more general phenomenon for how such materials grow. Lastly, this layer is likely highly defective and serves as a source of deleterious dopants. As is well-known for the epitaxial growth of $Bi_2Se_3$, the interface with $Al_2O_3$ is strongly n-type doped, which is electrostatically transferred into the $Bi_2Se_3$ film and leads to occupied topological surface states and bulk conduction states[19,21]. This fills up the bands and pushes the Dirac point down well below the Fermi energy. Therefore, this layer may be the source of such doping so a deeper understanding of how to balance charge-doping effects with physical effects such as enhanced diffusion is critical going forward.

**(ii) Nucleation and growth modes**: Nucleation of 2D materials at step edges is quite unique and likely strongly affects the growth modes, as highlighted from the AFM data in Fig. 6. From RHEED, $Bi_2Se_3$ is known to exhibit step-flow growth at the temperatures used here[23]. Pure step-flow growth occurs when layers nucleate purely from step edges, schematically shown in Fig. 6(gi). Layer-by-layer growth occurs by nucleation that occurs randomly, and then these monolayer-thick islands coalesce to form a continuous layer (Fig. 6(gii)). The AFM data in Fig. 6 suggest there are distinct growth modes, one being step-flow-like that is driven by step-edge nucleation and occurs when the terrace widths are narrow relative to the mean-free-path of atoms on the surface (Fig. 6 (b,c)). For $Bi_2Se_3$ the growth initially proceeds from the steps, as shown in Fig. 6(giii). This contrasts with the ideal step-flow model, which suggests the nucleation occurs simultaneously along the steps from which the growth homogeneously proceeds away from the step (essentially in 1D) or might randomly nucleate along steps, but then coalesces while maintaining epitaxial registry. In contrast, $Bi_2Se_3$ grows with a triangular habit, which nucleates along the steps and grows outward, as seen in the islands in Fig. 6. These layers then coalesce with a random distance between growth fronts since there is a large mismatch between the in-plane lattice parameters. This randomness inevitably results in myriad defects where these fronts meet, which is not captured in the simple step-flow growth mode. Moreover, since there is a large mismatch in the unit cell of $Bi_2Se_3$ (10 Å) compared to the $Al_2O_3$ steps (2 Å), the growth occurs on the lower side of the step and proceeds away, yet leaves a second nucleation point on the top side which can nucleate a new growth front on the opposite side, (Fig. 6(giii)).

The second growth mode occurs for thick films and thin films with large terrace width (Fig. 6(a, d-f)). This mode is driven by random nucleation, like layer-by-layer growth, but growth from the nucleation site proceeds like step-flow. As the data shows, this second growth mode occurs for large thicknesses and large terrace widths where the effective density of steps reduces to the point where nucleation is random. Interestingly, the cross-sectional HAADF-STEM clearly shows that the layers eventually overgrow the step-edges. This novel process drives this transition to this second growth mode due to the elimination of nucleation control, which also favors equal APT formation. This transition happens by exchanging the higher energy associated with the step-edge defect for the lower energy associated with shear straining the layer over the steps. The overlaid layer no longer acts as a nucleation site, thus changing the growth mode. Overlapping of the layer should occur randomly, and therefore it is expected that the films will fully become twinned at larger thicknesses. The parameters that are critical for triggering the overlaying of 2D layers remain an open question for future work, although it should strongly depend on temperature, growth rate, step-edge height, local bonding geometries, etc. Altogether, these findings are broadly applicable to 2D materials growth in general and are interesting points of future work for theoretical modeling. This will help understand and disentangle the fundamental origins of the dependence of the growth mode on growth parameters, which will enable growing higher-quality materials.

**(iii) Defects**: As highlighted in this work, we have been able to stabilize a single APT of $Bi_2Se_3$ using step-terminated $Al_2O_3$ with a 3° miscut. There is commonality with previous reports on InP(111) as well as other surfaces, given that the surfaces on InP(111) are rough with dense vertical steps, as, without a doubt, is $Bi_2Se_3$ on 3° miscut $Al_2O_3$. These aspects highlight that the solution found here may not be a complete success since the goal is to have $Bi_2Se_3$ with minimal defects. However, having defect sites that are spaced on average of 5 nm will still result in a net high-level of disorder and a low carrier mobility. Therefore, *the critical goal going forward will be to understand how to reduce APT on substrates with the largest terrace width, opening the path to preserving a high carrier mobility.*

Lastly, the novel step-edge-defects in the topological insulator may not be deleterious, but, rather, might be interesting for finding new types of edge states. It has been shown that novel Rashba states form at QL step edges in $Bi_2Se_3$[48] as well as locally strained regions[49]. The vertical defects shown here offer a materials route to explore and discover new physics at these boundaries. Interestingly, HAADF-STEM shows that there is clear and well-defined inter-QL bonding, and how this affects the emergent 1D states that arise at these boundaries is of significant interest. Compared to the QL step edges, the height of these step-edge defects is determined by the substrate used, which offers additional routes to tailor the properties since step-termination is well-established for a wide variety of substrates. To achieve this there are many additional questions that arise regarding the fundamental aspects that dictate the structures at the step-edges. As shown in the theoretical models in Fig. 2 and experimentally in Fig. 5(d), this includes how the bonds arrange at and across the layers and shear strain effects that arise due to the overgrowth. These will have profound impacts on the resulting states that emerge at these 1D boundaries.

**Conclusion**

To conclude, we have shown that step-terminated $Al_2O_3$ substrates can be used to effectively reduce twins in the prototypical topological insulator $Bi_2Se_3$ and more broadly 2D materials. Using first principles models, we showed that the 2 Å steps in $Al_2O_3$ were effective at introducing an energy barrier that selects a particular orientation of $Bi_2Se_3$. This was found to be consistent with macroscale x-ray diffraction measurements, where temperature and terrace width (miscut) point to the growth being nucleation-controlled, as the mean-free-path should be of the order of the terrace width. However, detailed structural HAADF-STEM images highlighted that there are unique aspects that create challenges to stabilize a single twin. Specifically, stabilizing pure step-flow growth is challenging in this system due to nuances at the step-edges, given the incommensurate step heights of $Al_2O_3$ and $Bi_2Se_3$, as well as $Bi_2Se_3$ displaying a proclivity to overgrow the steps, which nullifies the step as a nucleation center. When taken together, this highlights that there are many open theoretical and experimental questions regarding how these aspects can be used to connect parameters controllable at the growth stage, such as temperature, miscut, step height, and growth rate, with probabilities for nucleation of a single orientation. To achieve this, many aspects require clarification: nucleation and growth arising at steps of incommensurate lattice parameters, the energetics for layers overlaying the steps, and the role of the interface layer. This should motivate extending theoretical models such as the classical Burton-Cabrera-Frank (BCF) theory[50,51] and additional experiments to quantify and measure these key aspects. This basic insight can then be used to realize higher-quality materials, as well as novel platforms for designing and realizing new physics. Bond-geometry-control of 2D materials with strong spin-orbit coupling that is confined to 1D steps, moreover, can give rise to new topological states,[48,49] and this can offer new routes to tailoring these properties. Altogether, the work here shows that it is not only possible to achieve the growth of a single twin of $Bi_2Se_3$ but also highlights critical aspects that should generalize to other 2D layered materials. Future studies motivated by this work should not only result in higher quality materials but also in the realization of new physics that emerges due to materials-physics codesign.

**Data Availability**

The data supporting this study's findings are available from the corresponding author upon reasonable request.

**Acknowledgments**

This work was supported by the U. S. Department of Energy (DOE), Office of Science, Basic Energy Sciences (BES), Materials Sciences and Engineering Division (Growth and X-ray diffraction). The National Quantum Information Science Research Centers, Quantum Science Center, and the NNSA's Laboratory Directed Research and Development Program at Los Alamos National Laboratory (assistance in growth). Los Alamos National Laboratory is managed by Triad National Security, LLC for the U.S. Department of Energy's NNSA, under contract 89233218CNA00000 (Part of the X-ray diffraction). Y.C. acknowledges the support from the U.S. Department of Energy, Office of Science, Basic Energy Sciences, Materials Science and Engineering Division through the Early Career Research Program (X-ray diffraction analysis). The first principles calculations done at the University of Alabama at Birmingham is supported by the National Science Foundation under Grant No. OIA-2229498, and the ORAU Ralph E. Powe Junior Faculty Enhancement Award. This research used resources of the National Energy Research Scientific Computing Center (NERSC), a Department of Energy Office of Science User Facility using NERSC award CNMS2023-A-01934. The authors also gratefully acknowledge the resources provided by the University of Alabama at Birmingham IT-Research Computing group for high performance computing (HPC) support and CPU time on the Cheaha compute cluster. G.A.V.-L. and D.R.H. acknowledge support for electron microscopy through startup funds from the Penn State Eberly College of Science, Department of Chemistry, College of Earth and Mineral Sciences, Department of Materials Science and Engineering, and Materials Research Institute. G.A.V.-L. and J.B. acknowledges support from the DOE Office of Science, Office of Workforce Development for Teachers and Scientists, Office of Science Graduate Student Research (SCGSR) program. The SCGSR program is administered by the Oak Ridge Institute for Science and Education (ORISE) for the DOE. ORISE is managed by ORAU under contract number DE- SC0014664. This STEM specimens were prepared using Electron Microscopy Facility of the Center for Functional Nanomaterials (CFN), which is a U.S. Department of Energy Office of Science User Facility, at Brookhaven National Laboratory under Contract No. DE-SC0012704, and the STEM and AFM experiments were conducted as part of a user project at the Center for Nanophase Materials Sciences (CNMS), which is a US Department of Energy, Office of Science User Facility at Oak Ridge National Laboratory.

**References**

[1] M. Z. Hasan, C. L. Kane, *Reviews of Modern Physics* **2010**, *82*, 3045.

[2] H. Zhang, C.-X. Liu, X.-L. Qi, X. Dai, Z. Fang, S.-C. Zhang, *Nat Phys* **2009**, *5*, 438.

[3] Y. Xia, D. Qian, D. Hsieh, L. Wray, A. Pal, H. Lin, A. Bansil, D. Grauer, Y. S. Hor, R. J. Cava, M. Z. Hasan, *Nat Phys* **2009**, *5*, 398.

[4] J. P. Heremans, R. J. Cava, N. Samarth, *Nature Reviews Materials* **2017**, *2*, 1.

[5] N. P. Armitage, E. J. Mele, A. Vishwanath, *Reviews of Modern Physics* **2018**, *90*, 015001.

[6] Z. K. Liu, B. Zhou, Y. Zhang, Z. J. Wang, H. M. Weng, D. Prabhakaran, S.-K. Mo, Z. X. Shen, Z. Fang, X. Dai, Z. Hussain, Y. L. Chen, *Science* **2014**, *343*, 864.

[7] M. Neupane, S.-Y. Xu, R. Sankar, N. Alidoust, G. Bian, C. Liu, I. Belopolski, T.-R. Chang, H.-T. Jeng, H. Lin, A. Bansil, F. Chou, M. Z. Hasan, *Nat Commun* **2014**, *5*, 3786.

[8] Z. K. Liu, J. Jiang, B. Zhou, Z. J. Wang, Y. Zhang, H. M. Weng, D. Prabhakaran, S.-K. Mo, H. Peng, P. Dudin, T. Kim, M. Hoesch, Z. Fang, X. Dai, Z. X. Shen, D. L. Feng, Z. Hussain, Y. L. Chen, *Nature materials* **2014**, *13*, 677.

[9] S.-Y. Xu, I. Belopolski, N. Alidoust, M. Neupane, G. Bian, C. Zhang, R. Sankar, G. Chang, Z. Yuan, C.-C. Lee, S.-M. Huang, H. Zheng, J. Ma, D. S. Sanchez, B. Wang, A. Bansil, F. Chou, P. P. Shibayev, H. Lin, S. Jia, M. Z. Hasan, *Science* **2015**, *349*, 613.

[10] B. Q. Lv, H. M. Weng, B. B. Fu, X. P. Wang, H. Miao, J. Ma, P. Richard, X. C. Huang, L. X. Zhao, G. F. Chen, Z. Fang, X. Dai, T. Qian, H. Ding, *Physical Review X* **2015**, *5*, 031013.

[11] Q. Lu, P. V. S. Reddy, H. Jeon, A. R. Mazza, M. Brahlek, W. Wu, S. A. Yang, J. Cook, C. Conner, X. Zhang, A. Chakraborty, Y.-T. Yao, H.-J. Tien, C.-H. Tseng, P.-Y. Yang, S.-W. Lien, H. Lin, T.-C. Chiang, G. Vignale, A.-P. Li, T.-R. Chang, R. G. Moore, G. Bian, *Nat Commun* **2024**, *15*, 6001.

[12] D. J. P. de Sousa, S. Lee, Q. Lu, R. G. Moore, M. Brahlek, J.-P. Wang, G. Bian, T. Low, *Phys. Rev. Lett.* **2024**, *133*, 146605.

[13] A. F. May, J. Yan, M. A. McGuire, *Journal of Applied Physics* **2020**, *128*, 51101.

[14] M. Brahlek, J. Lapano, J. S. Lee, *Journal of Applied Physics* **2020**, *128*, 210902.

[15] A. K. Geim, I. V. Grigorieva, *Nature* **2013**, *499*, 419.

[16] M. Brahlek, *Advanced Materials* **2020**, *32*, 2005698.

[17] C.-Z. Chang, C.-X. Liu, A. H. MacDonald, *Reviews of Modern Physics* **2023**, *95*, 011002.

[18] D. Reifsnyder Hickey, K. A. Mkhoyan, *APL Materials* **2020**, *8*, 070902.

[19] M. Brahlek, N. Koirala, N. Bansal, S. Oh, *Solid State Communications* **2014**, *215*, 54.

[20] N. Koirala, M. Brahlek, M. Salehi, L. Wu, J. Dai, J. Waugh, T. Nummy, M.-G. Han, J. Moon, Y. Zhu, D. Dessau, W. Wu, N. P. Armitage, S. Oh, *Nano Letters* **2015**, *15*, 8245.

[21] M. Brahlek, N. Koirala, M. Salehi, N. Bansal, S. Oh, *Physical Review Letters* **2014**, *113*, 026801.

[22] D. Reifsnyder Hickey, R. J. Wu, J. S. Lee, J. G. Azadani, R. Grassi, M. DC, J.-P. Wang, T. Low, N. Samarth, K. A. Mkhoyan, *Phys. Rev. Mater.* **2020**, *4*, 011201.

[23] N. Bansal, Y. S. Kim, E. Edrey, M. Brahlek, Y. Horibe, K. Iida, M. Tanimura, G.-H. Li, T. Feng, H.-D. Lee, T. Gustafsson, E. Andrei, S. Oh, *Thin Solid Films* **2011**, *520*, 224.

[24] A. Richardella, A. Kandala, J. S. Lee, N. Samarth, *APL Materials* **2015**, *3*, 083303.

[25] X. Guo, Z. J. Xu, H. C. Liu, B. Zhao, X. Q. Dai, H. T. He, J. N. Wang, H. J. Liu, W. K. Ho, M. H. Xie, *Applied Physics Letters* **2013**, *102*, 151604.
[26] N. V. Tarakina, S. Schreyeck, M. Luysberg, S. Grauer, C. Schumacher, G. Karczewski, K. Brunner, C. Gould, H. Buhmann, R. E. Dunin-Borkowski, L. W. Molenkamp, *Advanced Materials Interfaces* **2014**, *1*, 1400134.
[27] S. Schreyeck, N. V. Tarakina, G. Karczewski, C. Schumacher, T. Borzenko, C. Brüne, H. Buhmann, C. Gould, K. Brunner, L. W. Molenkamp, *Applied Physics Letters* **2013**, *102*, 041914.
[28] X. Yao, J. Moon, S.-W. Cheong, S. Oh, *Nano Res.* **2020**, *13*, 2541.
[29] X. Yao, B. Gao, M. G. Han, D. Jain, J. Moon, J. W. Kim, Y. Zhu, S. W. Cheong, S. Oh, *Nano Letters* **2019**, *19*, 4567.
[30] M. Salehi, H. Shapourian, I. T. Rosen, M. Han, J. Moon, P. Shibayev, D. Jain, D. Goldhaber-Gordon, S. Oh, *Advanced Materials* **2019**, *31*, 1901091.
[31] K. S. Wickramasinghe, C. Forrester, M. C. Tamargo, *Crystals* **2023**, *13*, 677.
[32] M. Brahlek, N. Bansal, N. Koirala, S.-Y. Xu, M. Neupane, C. Liu, M. Z. Hasan, S. Oh, *Physical Review Letters* **2012**, *109*, 186403.
[33] L. Wu, M. Brahlek, R. Valdés Aguilar, A. V. Stier, C. M. Morris, Y. Lubashevsky, L. S. Bilbro, N. Bansal, S. Oh, N. P. Armitage, *Nature Physics* **2013**, *9*, 410.
[34] M. Salehi, H. Shapourian, N. Koirala, M. J. Brahlek, J. Moon, S. Oh, *Nano Letters* **2016**, *16*, 5528.
[35] M. Brahlek, N. Koirala, J. Liu, T. I. Yusufaly, M. Salehi, M.-G. Han, Y. Zhu, D. Vanderbilt, S. Oh, *Physical Review B* **2016**, *93*, 125416.
[36] X. Liu, D. J. Smith, J. Fan, Y.-H. Zhang, H. Cao, Y. P. Chen, J. Leiner, B. J. Kirby, M. Dobrowolska, J. K. Furdyna, *Applied Physics Letters* **2011**, *99*, 171903.
[37] A. Koma, *Thin Solid Films* **1992**, *216*, 72.
[38] G. Kresse, J. Furthmüller, *Computational Materials Science* **1996**, *6*, 15.
[39] G. Kresse, J. Furthmüller, *Physical Review B* **1996**, *54*, 11169.
[40] G. Kresse, D. Joubert, *Physical Review B* **1999**, *59*, 1758.
[41] J. P. Perdew, K. Burke, M. Ernzerhof, *Physical Review Letters* **1996**, *77*, 3865.
[42] H. J. Monkhorst, J. D. Pack, *Physical Review B* **1976**, *13*, 5188.
[43] J. R. Young, M. Chilcote, M. Barone, J. Xu, J. Katoch, Y. K. Luo, S. Mueller, T. J. Asel, S. K. Fullerton-Shirey, R. Kawakami, J. A. Gupta, L. J. Brillson, E. Johnston-Halperin, *Applied Physics Letters* **2017**, *110*, 263103.
[44] A. R. Mazza, A. Miettinen, A. A. Daykin, X. He, T. R. Charlton, M. Conrad, S. Guha, Q. Lu, G. Bian, E. H. Conrad, P. F. Miceli, *Nanoscale* **2019**, *11*, 14434.
[45] H. Yi, L.-H. Hu, Y. Wang, R. Xiao, J. Cai, D. R. Hickey, C. Dong, Y.-F. Zhao, L.-J. Zhou, R. Zhang, A. R. Richardella, N. Alem, J. A. Robinson, M. H. W. Chan, X. Xu, N. Samarth, C.-X. Liu, C.-Z. Chang, *Nature Materials* **2022**, *21*, 1366.
[46] M. Xu, K. Ye, I. Sadeghi, R. Jaramillo, J. M. LeBeau, *J. Vac. Sci. Technol. A* **2024**, *42*, 063207.
[47] N. Hagopian, T. Jung, Z. LaDuca, T. Samanta, A. Sirohi, K. Su, C. Dong, Q. T. Campbell, M. S. Arnold, J. A. Robinson, J. K. Kawasaki, P. M. Volyes, *Microanal* **2025**, *31*, ozaf048.576.
[48] W. Ko, S.-H. Kang, J. Lapano, H. Chang, J. Teeter, H. Jeon, Q. Lu, A.-H. Chen, M. Brahlek, M. Yoon, R. G. Moore, A.-P. Li, *ACS Nano* **2024**, *18*, 18405.
[49] Y. Liu, Y. Y. Li, S. Rajput, D. Gilks, L. Lari, P. L. Galindo, M. Weinert, V. K. Lazarov, L. Li, *Nature Physics* **2014**, *10*, 294.

[50] W. K. Burton, N. Cabrera, F. C. Frank, N. F. Mott, *Philosophical Transactions of the Royal Society of London. Series A, Mathematical and Physical Sciences* **1951**, *243*, 299.

[51] D. P. Woodruff, *Philosophical Transactions of the Royal Society A: Mathematical, Physical and Engineering Sciences* **2015**, *373*, 20140230.

**Figure 1:**

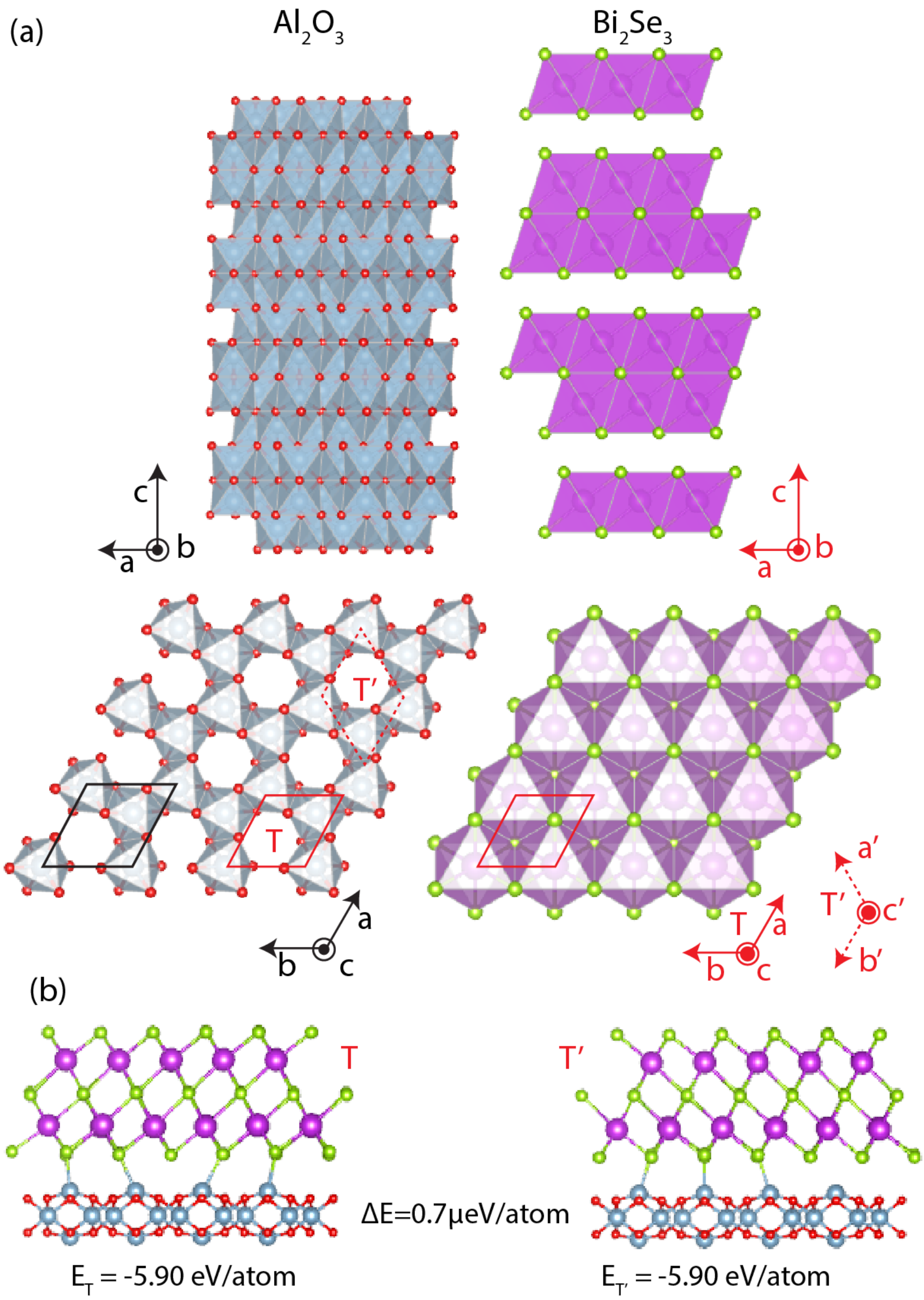

**Fig. 1** (a) Sideview and topview of the crystal structures for $Al_2O_3$, left, and $Bi_2Se_3$, right. The black coordinate system is that of $Al_2O_3$, and the red is that for $Bi_2Se_3$, where *T* and *T'* indicate possible orientations of $Bi_2Se_3$ on $Al_2O_3$. (b) Total energy per atom as calculated by density functional theory for *T* and *T'* orientations of $Bi_2Se_3$ on $Al_2O_3$.

**Figure 2**

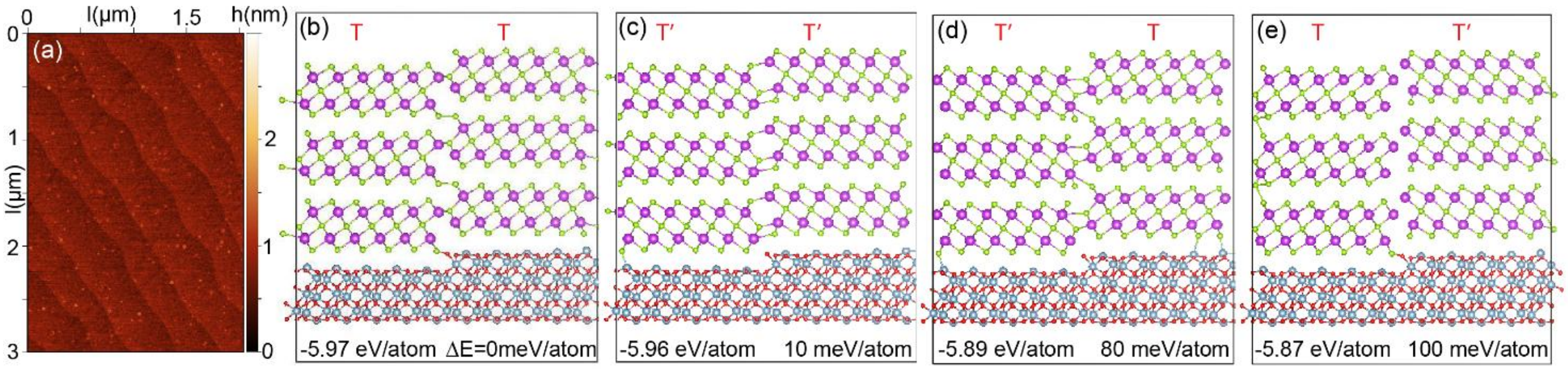

**Fig. 2.** (a) AFM image of a step-terminated $Al_2O_3$ substrate with miscut of 0.05°. (b-e) Total energy calculations for various alignments of twins at a step-edge of $Al_2O_3$. Here *T*-*T* (b) *T'*-*T'*(c), *T'*-*T*(d), and *T*-*T'* (e) with total energy per atom indicated in the lower left and the difference of energy ($\Delta E$) per atom relative to the *T*-*T* indicated in the lower right.

**Figure 3**

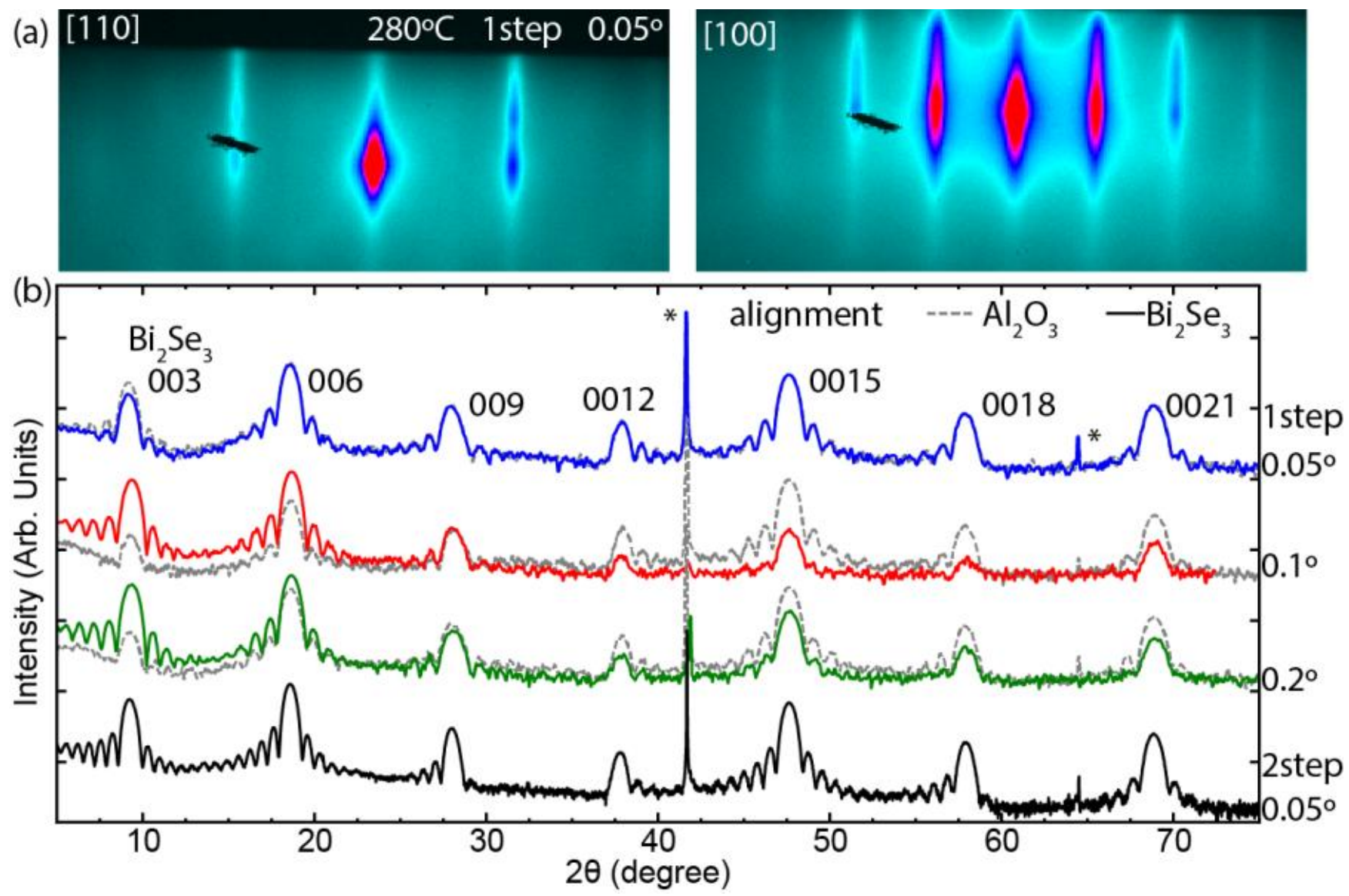

**Fig. 3.** (a) RHEED images along the [110] (left) and the [100] directions. (b) X-ray diffraction 2$\theta$-$\theta$ scans for $Bi_2Se_3$ grown on $Al_2O_3$ with various miscuts and 1-step and 2-step. For the solid-colored curves the offset in $\theta$ was aligned to maximize signal of the $Bi_2Se_3$ 006 and for the gray dashed curves $\theta$ was aligned to maximize the $Al_2O_3$ 006, which highlights the slightly misaligned 00*L* lattice planes.

**Figure 4**

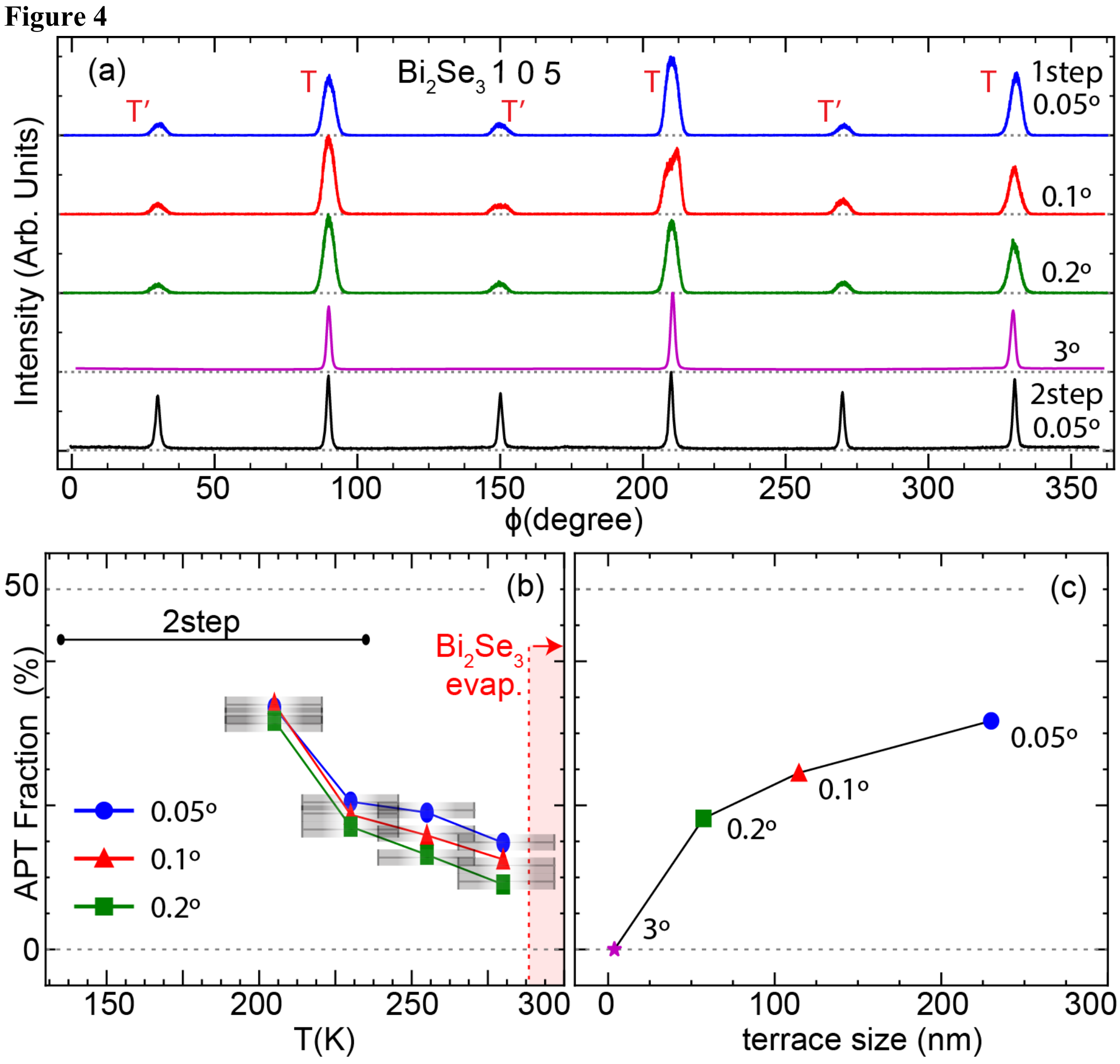

**Fig. 4** (a) Azimuthal scans about $\phi$ of the $Bi_2Se_3$ 105 peak various miscuts and 1 step and 2 step, since the 105 peak is three fold, the peaks at $\phi$=30, 150, 270° are labeled *T'* and those at 90, 210, 330° are *T*. (b) The relative ratio of the intensities of *T* and *T'* peaks (APT fraction) for various miscuts and 1 step and 2 step growths versus temperature, as labeled. The error bars were taken to be ±15 °C, and the dashed vertical line at ~280 °C indicates the point where $Bi_2Se_3$ evaporates. (c) APT fraction for films grown at 225 °C versus terrace size inferred from miscut.

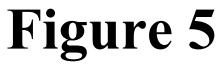

**Figure 5**

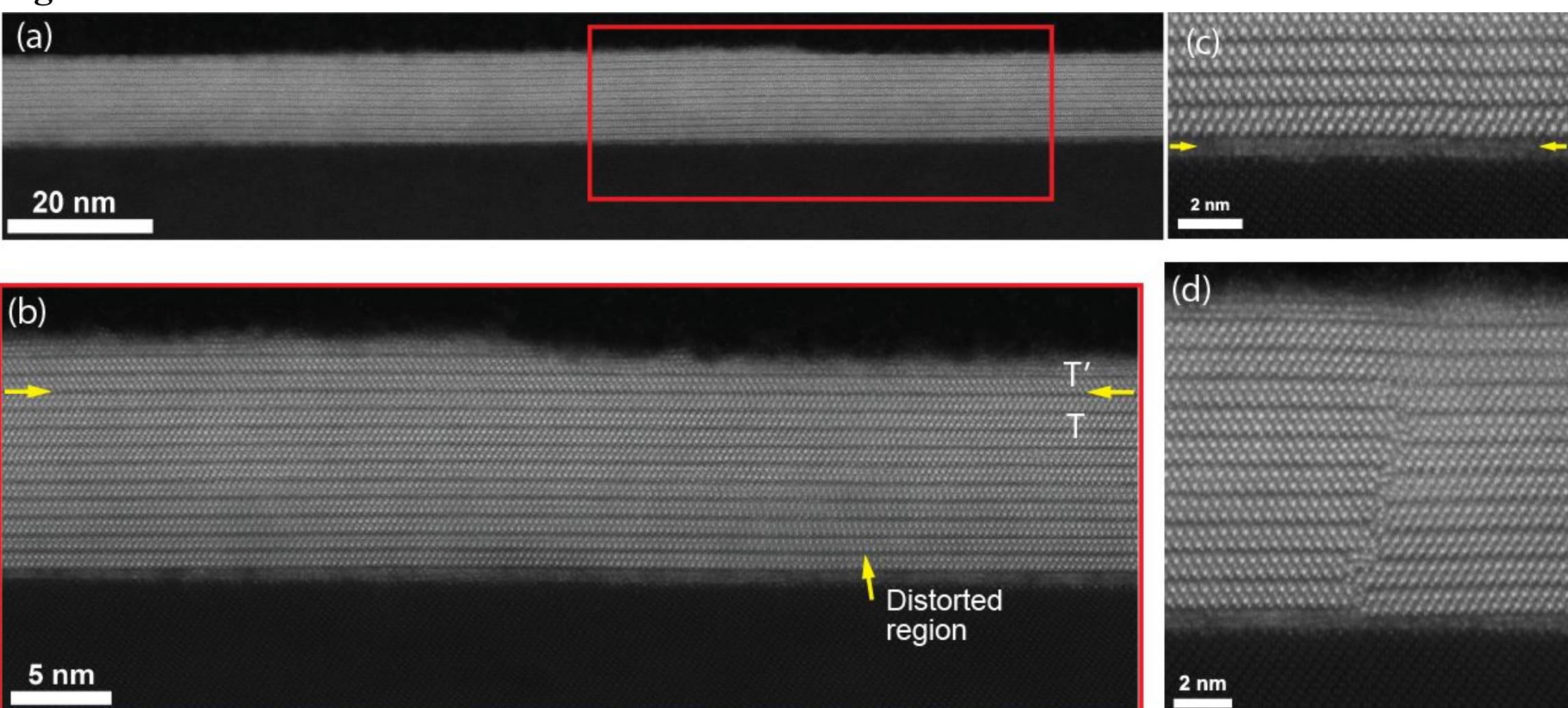

**Fig. 5.** (a) Low-magnification HAADF-STEM image of $Bi_2Se_3$ film. (b) Magnified region in the red box of figure (a). (c) HAADF-STEM image of the interfacial layer between the $Bi_2Se_3$ film and the $Al_2O_3$ substrate. (d) HAADF-STEM image of a nucleation point at a step edge being overlaid by a $Bi_2Se_3$ QL.

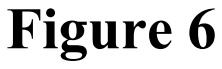

**Figure 6**

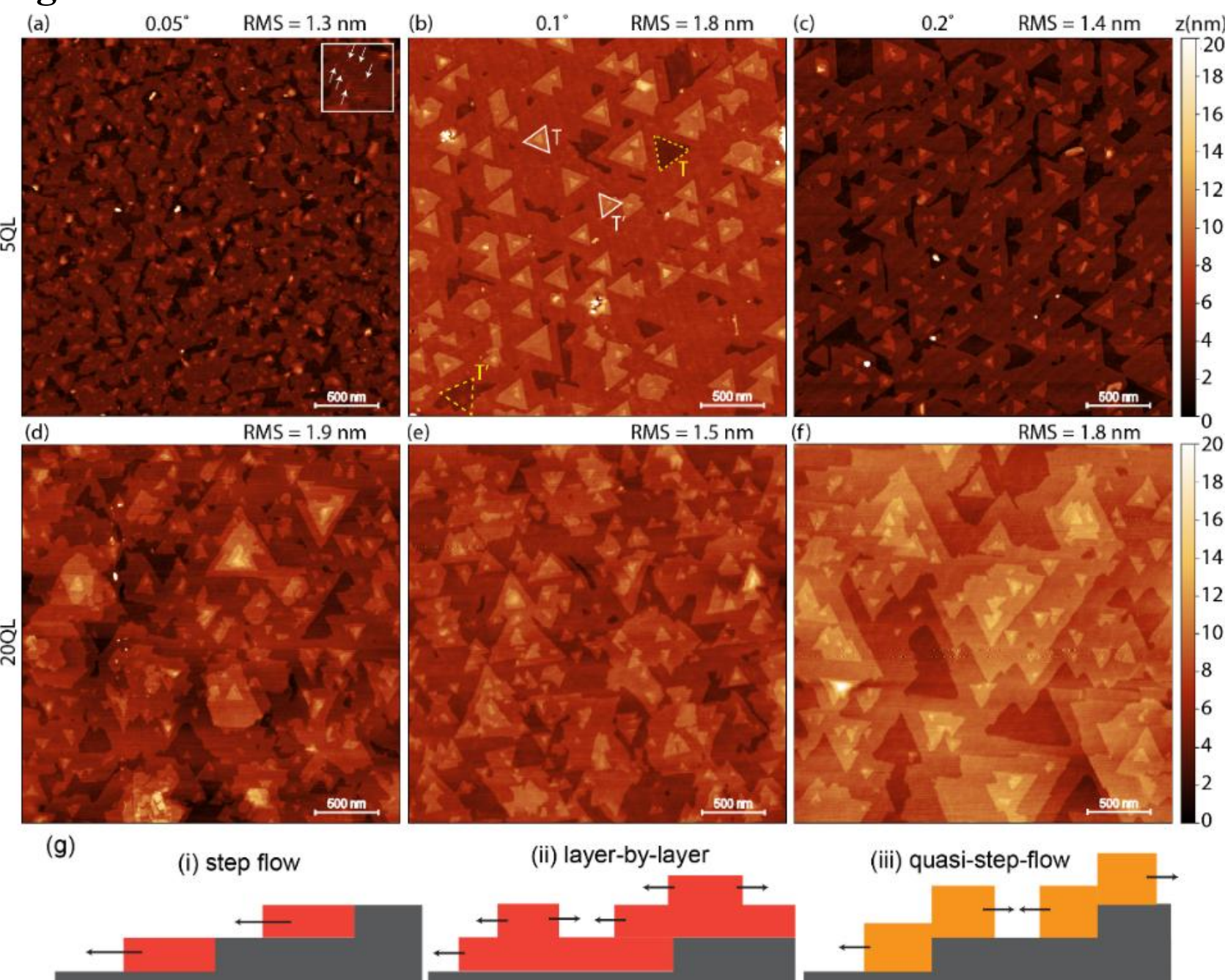

**Fig. 6.** AFM images for samples grown at 260 °C with thicknesses of 5 QL (a, b, c) and 20 QL (d, e, f) and miscut of $<0.05°$ (a, d), 0.1° (b, e), and 0.2° (c, f). Lateral size of each image is 3μm × 3μm. (g) Schematics of growth modes. (i-iii) Typical representation of step-flow (i) and layer-by-layer growth (ii) of a heterogeneous system with similar c-axis lattice parameters, and (iii) quasi-step-flow growth in a system with incommensurate c-axis lattice parameters.

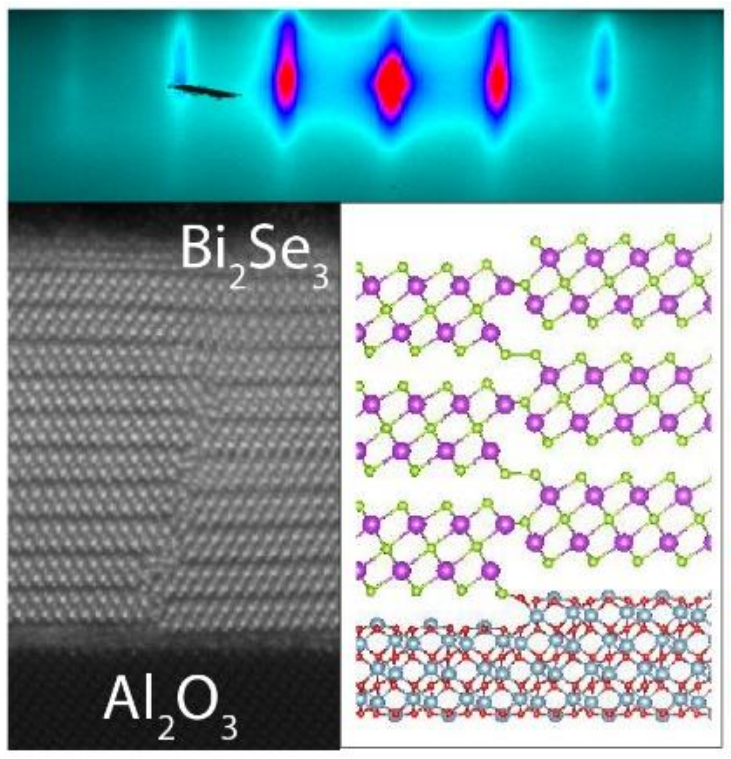

Overcoming a longstanding problem, step-terminated $Al_2O_3$ substrates enable controlled growth of high-quality $Bi_2Se_3$ films by suppressing antiphase twin defects. Combining experiments and theory, this work reveals how atomic step edges guide nucleation to produce single-domain topological insulator films, offering new insights for defect control and improved material performance in quantum and electronic applications.

# Supplementary Information

# Nucleation and Antiphase Twin Control in $Bi_2Se_3$ via Step-Terminated $Al_2O_3$ Substrates

Alessandro R. Mazza[1#], Jia Shi[2], Gabriel A. Vázquez-Lizardi[3], Sangsoo Kim[4], Jackson Bentley[4], An-Hsi Chen[4], Kim Kisslinger[5], Debarghya Mallick[4], Qiangsheng Lu[4], T. Zac Ward[6], Vitalii Starchenko[7], Nicholas Cucciniello[8,9], Robert G. Moore[4], Gyula Eres[4], Yue Cao[10], Debangshu Mukherjee[11], Liam Collins[6], Christopher Nelson[6], Danielle Reifsnyder Hickey[3,12,13], Fei Xue[2], Matthew Brahlek[4*]

[1]Materials Science and Technology Division, Los Alamos National Laboratory, Los Alamos, NM 87545, USA
[2]Department of Physics, University of Alabama at Birmingham, Birmingham, AL, 35294, USA
[3]Department of Chemistry, Pennsylvania State University, University Park, PA 16802, USA
[4]Materials Science and Technology Division, Oak Ridge National Laboratory, Oak Ridge, TN, 37831, USA
[5]Center for Functional Nanomaterials, Brookhaven National Laboratory, Upton, New York 11973, USA
[6]Center for Nanophase Materials Science, Oak Ridge National Laboratory, Oak Ridge, TN, 37831, USA
[7]Chemical Sciences Directorate, Oak Ridge National Laboratory, Oak Ridge, TN, 37831, USA
[8]Center for Integrated Nanotechnologies, Los Alamos National Laboratory, Los Alamos, NM 87545, USA
[9]Department of Materials Design and Innovation, University at Buffalo – The State University of New York, Buffalo, New York 14260, USA
[10]Materials Science Division, Argonne National Laboratory, Lemont, IL 60439, USA
[11]Computing and Computational Sciences Directorate, Oak Ridge National Laboratory, Oak Ridge, TN, 37831, USA
[12]Department of Materials Science and Engineering, Pennsylvania State University, University Park, PA 16802, USA
[13]Materials Research Institute, Pennsylvania State University, University Park, PA 16802, USA

Correspondence should be addressed to #armazza@lanl.gov, *brahlekm@ornl.gov

Experimental details

Atomic Force Microscopy (AFM) Characterization

Surface morphology was characterized using a DriveAFM system (Nanosurf AG, Switzerland) operated in WaveMode. A setpoint force of 5 nN was maintained during imaging to ensure stable tip–sample contact while minimizing tip-induced deformation. We employed NSC36 silicon cantilevers (MikroMasch) with the following nominal specifications: resonance frequency ~75 kHz, force constant ~2 N/m, and tip radius <10 nm. All AFM data were processed using Gwyddion (version 2.66). To remove background artifacts and scanner bow, each image was first-order flattened prior to further statistical analysis. Roughness parameters including RMS roughness (Sq), mean roughness (Sa), skewness (Ssk), were extracted from the flattened topography maps. Representative images are shown in the main text, and full datasets are available upon request.

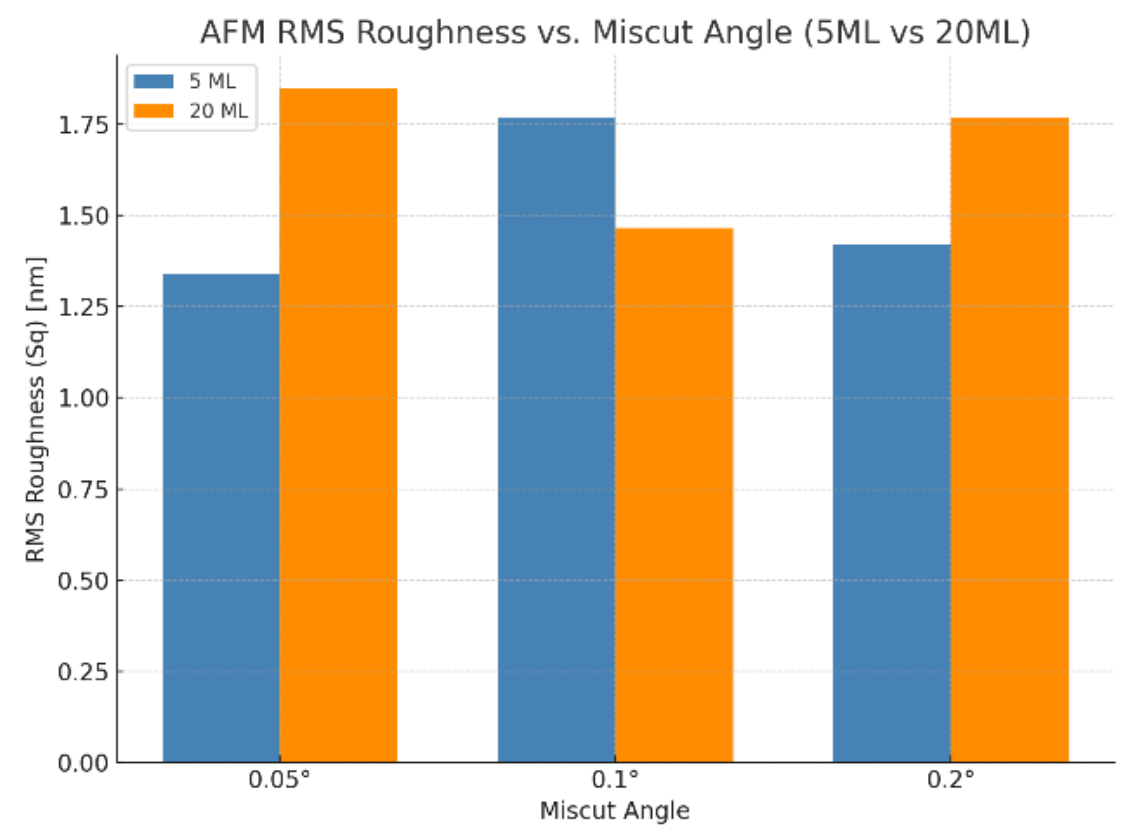

| Miscut | 5ML RMS (Sq) [nm] | 20ML RMS (Sq) [nm] |
|---|---|---|
| 0.05 | 1.33951 | 1.84729 |
| 0.1 | 1.76813 | 1.46424 |
| 0.2 | 1.4208 | 1.76806 |